\documentclass[aps,prl,showpacs,superscriptaddress,twocolumn]{revtex4}
\usepackage{amsmath}
\usepackage{amsfonts}
\usepackage{graphicx}
\usepackage[normalem]{ulem}
\usepackage{color}

\begin{document}

\author{P\'{e}ter F\"{o}ldi}
\affiliation{Department of Theoretical Physics, University of Szeged, Tisza Lajos k\"{o}r\'{u}t 84, H-6720 Szeged, Hungary}
\email{foldi@physx.u-szeged.hu}
\author{Istv\'an M\'arton}
\affiliation{Wigner Research Center for Physics, Konkoly-Thege Mikl\'{o}s \'{u}t 29-33, H-1121 Budapest, Hungary}
\affiliation{Department of Experimental Physics, University of P\'{e}cs, Ifjus\'{a}g \'{u}tja 6, H-7624 P\'{e}cs, Hungary}
\author{Nikolett N\'emet}
\affiliation{Wigner Research Center for Physics, Konkoly-Thege Mikl\'{o}s \'{u}t 29-33, H-1121 Budapest, Hungary}
\author{P\'eter Dombi}
\affiliation{Max-Planck-Institut f\"ur Quantenoptik, Hans-Kopfermann-Str. 1, D-85748 Garching, Germany}
\affiliation{Wigner Research Center for Physics, Konkoly-Thege Mikl\'{o}s \'{u}t 29-33, H-1121 Budapest, Hungary}

\title{Nanoscale optical waveform control of strong-field photoemission}

\begin{abstract}
Strong-field photoemission from metal nanostructures enabled fundamental discoveries recently. Here, we deliver theoretical demonstration of the electric field control of electrons in the closest nanoscale vicinity of plasmonic nanoparticles with the help of few-cycle laser waveforms. We analyze the effect of plasmonic resonance on photoemission properties and show that it is only off-resonant nanoparticles that can provide electron control on a true sub-fs timescale.
\end{abstract}


\maketitle

\bigskip
{\em Introduction}$\ \ $
The interaction of few-optical-cycle and/or mid-infrared femtosecond laser pulses with plasmonic thin films \cite{ID04,IE05,Dombi10,BG10,RI11}, metal or dielectric nanoparticles \cite{KO05,ZF11,DH13,GR12,NR13} and nanotips \cite{KS11,HS12,PS13} brought significant advances in recent years in understanding fundamental properties of ultrafast laser-solid interaction on the nanoscale. Among these discoveries are plasmonic strong-field photoemission \cite{ID04,IE05,Dombi10,BG10,RI11,DH13}, adaptive control of nano-optical near fields \cite{AB07}, quenching of electron quiver motion with mid-infrared laser pulses \cite{HS12} as well as control of above-threshold and strong-field photoemission in the vicinity of non-plasmonic nanotips with the waveform of few-cycle laser pulses \cite{KS11,PS13}.
\begin{figure}[htb]
\begin{center}
\includegraphics[width=8cm]{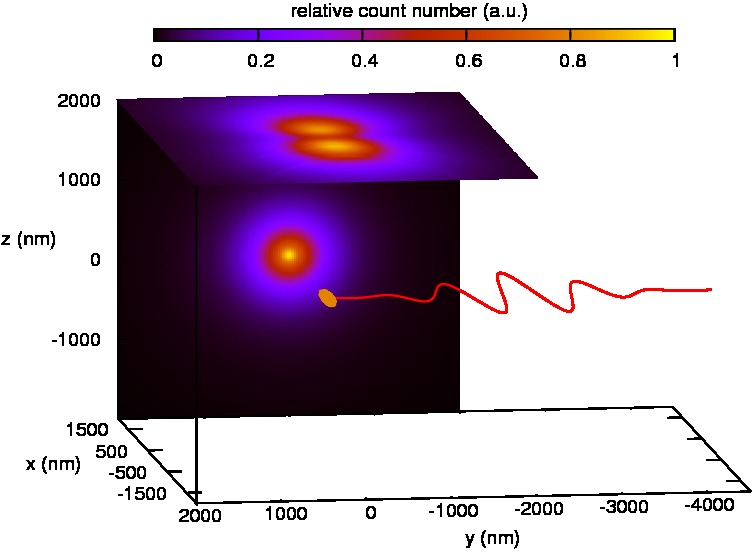}
\caption{A schematic view of the process we that we consider. The laser pulse (represented by the red wave) impinges on a nanoellipsoid (plotted by orange) exciting plasmonic oscillations, which lead to photoemission. The electrons are driven by the net electric field of the incident laser pulse and localized plasmons, and can be detected e.g., at the planes indicated where characteristic electron count distributions are also shown.
\label{geometryfig}}
\end{center}
\end{figure}

There are two fundamentally different approaches that are pursued to investigate these phenomena. The first one relies on electrochemically etched tungsten or gold nanotips that represent a single, well-defined nanoemitter, however, without any plasmonic resonances. Thus, field enhancement factors are low (only based on the tip effect) and the field geometry is strictly limited by the fixed tip shape resulting from the etching process. In contrast, resonant plasmonic nanoparticles can deliver field enhancement factors of several hundreds \cite{SB10}, moreover, electric field distribution on the nanoscale can be precisely controlled with the shape of the nanoparticle. These advantages were recently exploited in strong-field interaction studies where a clear correlation between plasmonic resonance and  nanoparticle photoemission spectra was demonstrated \cite{DH13}. However, related experiments were realized either with relatively long, multicycle laser pulses \cite{DH13} or with few-cycle pulses without carrier-envelope phase (i.e. optical waveform) control \cite{NR13},  Predictable steering of electrons on the nanoscale is only achievable if not only the envelope but also the waveform of the laser pulse is stabilized.
Therefore, here we analyze ultrafast photoemission and electron acceleration in the vicinity of plasmonic nanoparticles excited by carrier-envelope phase (CEP) stabilized, few-cycle laser pulses for the first time. In particular, the effect of plasmonic resonances on measurable photoelectron quantities is investigated and special attention is paid to resonant vs. off-resonant effects. As the spectral bandwidth of few-cycle laser pulses exceeds that of typical nanoparticle resonances, effective bandpass filtering exercised by plasmon coupling is expected to play a fundamental role. We will show how this limitation can be circumvented together with keeping considerable field enhancement factors. Our theoretical approach is partly analytic in determining electric field distribution in the vicinity of ellipsoidal nanoparticles and partly numerical in calculating classical trajectories of plasmonically photoemitted electrons.  The quantum nature of the photoemission process is taken into account by adapting a non-adiabatic tunneling ionization formula for this particular configuration \cite{YI01}. This way we can account for both low-intensity (multiphoton) and high-intensity (strong-field) processes at the same time. Note that although we present results corresponding to specific, typical parameter values, our qualitative conclusions are general.

\bigskip
{\em Model and results}$\ \ $
\begin{figure}[htb]
\begin{center}
\includegraphics[width=8cm]{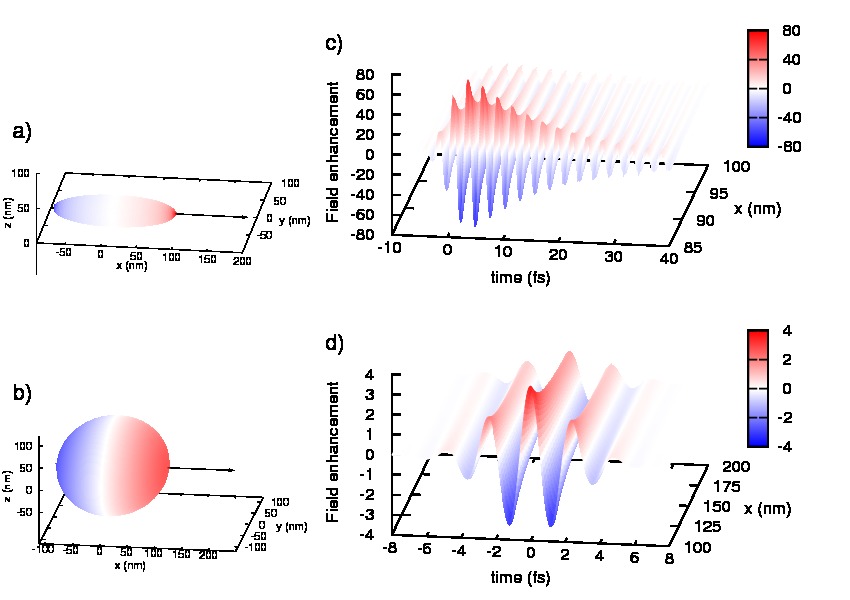}
\caption{Panels a) and b): the distribution of the normal component of the net electric field along the surface of the nanoparticles that we consider: a prolate ellipsoid $a=85, b=c=20 \ \mbox{nm},$ being resonant at the wavelength of $\lambda=800\ \mbox{nm},$ and an off-resonant sphere ($r=100 \ \mbox{nm}).$ These snapshots were taken at the maxima of the net electric field, the black arrows indicate the lines along which the spatial dependence of ${E}_x(x,t)$ is plotted in panels c) and d).
\label{awayfieldfig}}
\end{center}
\end{figure}
We consider an ellipsoid-shaped, nanometer-sized metallic particle interacting with a strong, pulsed laser field. The geometry of the problem -- corresponding to a possible experimental setup -- is summarized in Fig.~\ref{geometryfig}. The nanoellipsoid (yellow/gold) is centered at the origin, with its principal semi-axes (with lengths $c\leq b \leq a$) being parallel to the coordinate axes.
The incoming laser pulse (red oscillating curve in Fig.~\ref{geometryfig}) propagates along the $y$ axis, and its polarization is parallel to the $x$ axis. We assume that the time dependence of the laser field is given by
\begin{equation}
\mathbf{E}_L(t)=\hat{\mathbf{x}}\mathcal{E}_0\cos(\omega_0
t+\varphi_{\mathrm{CEP}}) \exp(-\frac{t^2}{2\tau^2}),
\label{pulse}
\end{equation}
where $\omega_0$ is the carrier frequency, and $\varphi_{\mathrm{CEP}}$ denotes the carrier-envelope phase (CEP) of the few-cycle laser pulse.
The parameters we use in the following correspond to existing sources in the near-infrared regime \cite{DA04,SS12}: $\omega_0=2.36\ \mbox{fs}^{-1}$ ($\lambda=800 \ \mbox{nm}$) and $\tau=2.3\ \mbox{fs}$ [corresponding to a full width at half maximum (FWHM) pulse length of $3.8\ \mbox{fs}$)].  The field amplitude $\mathcal{E}_0$ ranges from 1 to 10 GV/m.

Since the size of the nanoparticles are considerably below the wavelength of the laser field, one can use quasistatic approximation for obtaining the net electric field, which is a sum of the exciting laser pulse and the response of the nanoparticle, i.e., the plasmonic field \cite{BH83}:
\begin{equation}
\mathbf{E}(\mathbf{r},t)=\mathbf{E}_L(t)+\mathbf{E}_P(\mathbf{r},t).
\label{field}
\end{equation}
The shape of the nanoparticles allows us to calculate the space and time dependent plasmonic field analytically \cite{BH83,SS11}. For the sake of definiteness, we took material parameters for the dielectric function that correspond to Au \cite{E87,B05}. (Note that the particles are assumed to be surrounded by a dielectric material with relative permittivity of $\varepsilon=2.5,$ corresponding to a typical substrate material.) The results are visualized in Fig.~\ref{awayfieldfig} for the two different nanoparticles that we examine below: a resonant prolate ellipsoid and a nonresonant sphere. Considering the amplitude of the exciting laser pulse, we have chosen $\mathcal{E}_0=10$ and $1\ \mbox{GV/m}$ for the ellipsoid and the sphere, respectively. Due to field enhancement differences, the peak value of the net electric field is still higher for the resonant prolate geometry than the spherical one, but they are comparable. The net electric field is the strongest along the surface at $x=\pm a,$ i.e., where the particle has maximal extension in the direction parallel to the polarization of the incoming laser field, with maximum field enhancement factors of 80 (ellipsoid) and 4 (sphere).  According to the second column of panels in Fig.~\ref{awayfieldfig}, the plasmonic field decays fast as a function of the distance form the surface. Focusing on the time dependence, we can see that in case of resonance plasmonic oscillations are still observable several tens of femtoseconds after the disappearance of the exciting laser pulse. This is a crucial point, since CEP-dependent effects are expected to become negligible as the number of the optical cycles increases in a pulse. Therefore, in order to focus on the control of electrons by true few-cycle optical fields, in the following we concentrate on the nonresonant case. 
\begin{figure}[htb]
\begin{center}
\includegraphics[width=8cm]{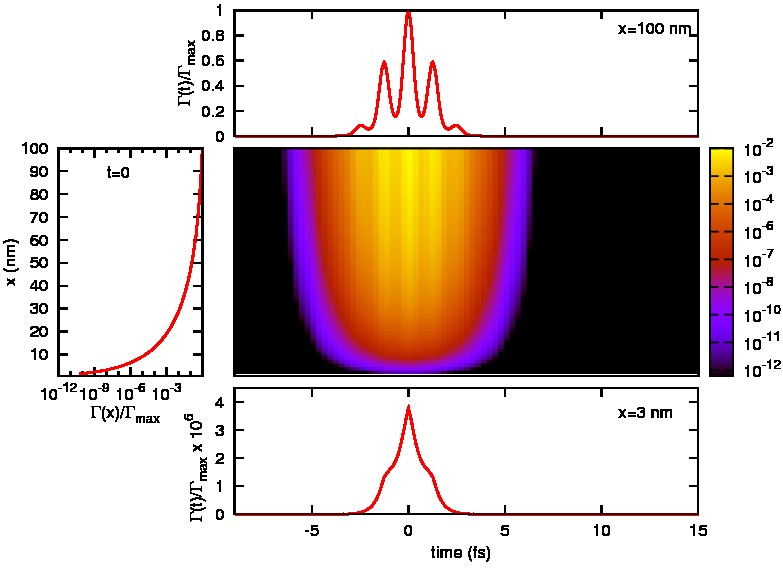}
\caption{Space and time dependence of the photoemission rate $\Gamma$ for a nonresonant sphere $a=b=c=100 \ \mbox{nm},$ with $\mathcal{E}_0=10\ \mbox{GV/m.}$ The central false-colored panel shows $\Gamma(\mathbf{r},t)$ along the $y=0$ line on the surface, while the side panels are cross sections at the indicated time instant and positions.
\label{excite2fig}}
\end{center}
\end{figure}

Having obtained the net electric field, we calculate the probability of photoemission per unit time and surface area. To this end, we implement the results of Ref.~\citenum{YI01}, where, based on quantum mechanical grounds, the authors developed a versatile formula for the photoionization rate $\Gamma(t)$ from a single atom. The expression is essentially an exponential function multiplied by a prefactor. The latter is obviously different for metal surfaces and atoms, but the saddle-point analysis (for more details, see also Ref.~\citenum{HI13}) leading to the exponential term is valid also for nanoparticles in the parameter range we consider. That is, although the overall electron yield may not be exact, the subcycle dependence is correct. In other words, using the results of Ref.~\citenum{YI01}, we can calculate the relative rates for different CEP values with the same envelope function [see Eq.~(\ref{pulse})] appropriately.

We assume that there is no photoemission (i.e., $\Gamma=0$) at a certain surface point when the local field points outward. This means that at a certain time instant photoelectrons emerge from only one of the semiellipsoids characterized by $x>0$ or $x<0.$  Focusing on points where $\Gamma(\mathbf{r},t)\neq 0,$  Fig.~\ref{excite2fig} shows the space and time dependence of the photoemission rate. As we can see, around $x=a,$ where the net electric field is maximal (cf.~Fig.~\ref{awayfieldfig}), $\Gamma$ is also maximal, and its time dependence adiabatically follows the oscillations of the local field \cite{YI01}. Around $x=0,$ however, the photoemission rate is by orders of magnitude lower, and its time dependence is completely determined by the envelope of the net electric field \cite{YI01} -- as expected in the case of multiphoton-induced photoemission.

\begin{figure}[htb]
\begin{center}
\includegraphics[width=8cm]{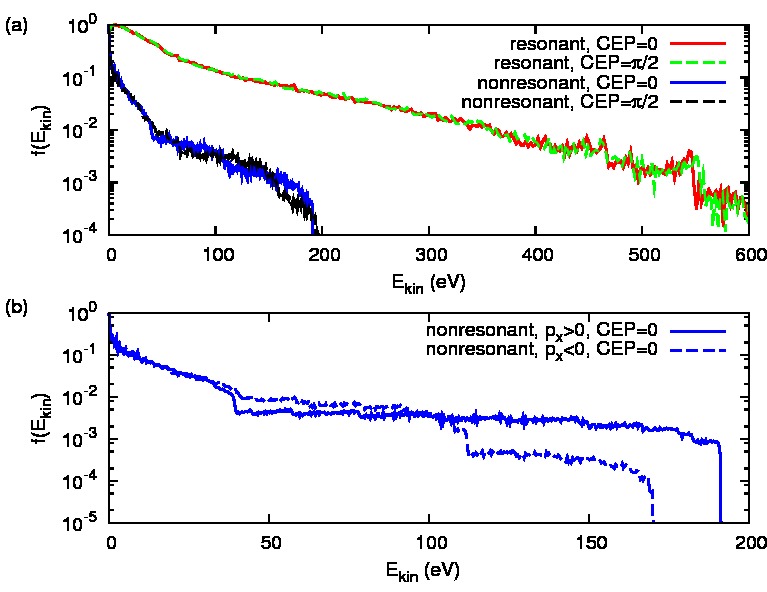}
\caption{Panel (a): Total photoelectron energy spectra for resonant and nonresonant nanoellipsoids. The peak electric field of the incident laser pulse is $\mathcal{E}_0=1\ \mbox{GV/m}$ for the red and green curves,  and $\mathcal{E}_0=10\ \mbox{GV/m}$ for the blue and black ones. CEP values are indicated in the legend. Panel (b): The energy spectra of photoelectrons with positive and negative (solid and dashed blue lines, respectively) final momentum in the $x$ direction (nonresonant case). Note that this panel corresponds to the case of opposite facing detectors.
\label{spect1fig}}
\end{center}
\end{figure}

Using the photoemission rate $\Gamma(\mathbf{r},t)$, we can calculate the probability weight that can be associated to an electron that emerges at the time instant $t$ and surface point $\mathbf{r}.$ After that, we assume that the net electric field results in a Lorentz force that accelerates negatively charged classical particles. We solve the corresponding equations of motion numerically, with a spacetime mesh of $20000$ initial surface points and time intervals of $0.05\ \mbox{fs}.$ Consistently, electron rescattering processes from the nanoparticle surface are also taken into account by classical, geometrical means. In the following, $f(X_0)$ denotes the relative likelihood of the appearance of a certain value $X_0$ of the physical quantity $X.$ The probability distribution (or rather density) functions $f(X)$ are normalized by setting their maxima to unity.

Let us start our statistical analysis with the total energy spectra $f(E_{kin})$ of the photoemitted electrons. Fig.~\ref{spect1fig}a) shows representative examples for both the resonant and the nonresonant cases. As we can see, although the amplitude of the incoming laser field is ten times larger in the nonresonant case, considerably higher energies occur for the resonant prolate ellipsoid, since the field enhancement factor is around $80$. Additionally,in accordance with the expectation that CEP-related effects are less important for longer, many-cycle pulses, we can see clear CEP dependence in the total photoelectron spectra only when the exciting field is offresonant (c.f.~Fig.~\ref{awayfieldfig}).   To demonstrate a higher degree of control by the laser waveform, we consider the spectra of electrons with positive and negative final momentum in the $x$ direction separately, i.e., spectra that can be measured by detectors that collect electrons on different sides of the nanoellipsoid. The result can be seen in Fig.~\ref{spect1fig} for cosine ($\varphi_{\mathrm{CEP}}=0$) and sine ($\varphi_{\mathrm{CEP}}=\pi/2$) pulses, with a CEP dependence being most pronounced for higher energies. In fact, there is a well defined energy range between $170$ and $190\ \mbox{eV}$, in which we can find a considerable number of electrons for the cosine pulse, while particles corresponding to these energies are practically absent for the sine pulse. The main reason for this effect is that the maximal electric field strength is higher in the cosine pulse (provided the envelope function is the same).
\begin{figure}[htb]
\begin{center}
\includegraphics[width=8cm]{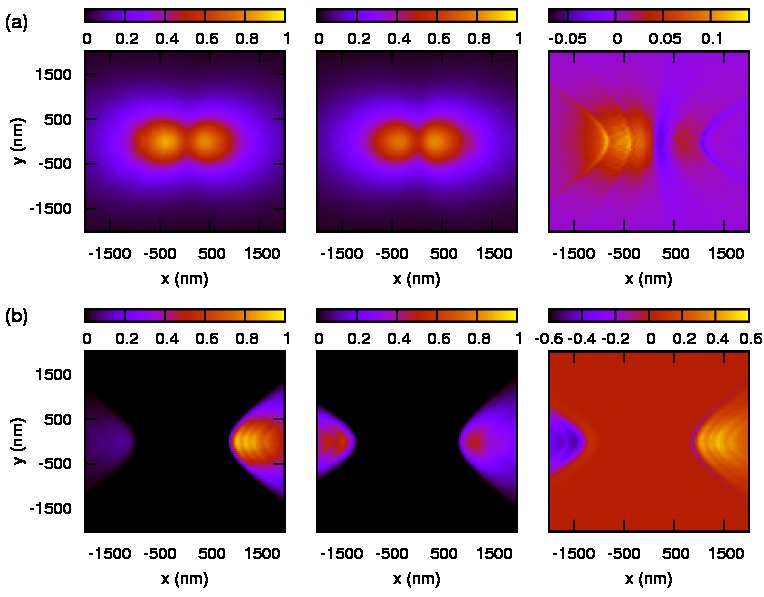}
\caption{Position distribution of the photoemitted electrons (relative count numbers) from off-resonant nanospheres at a possible detection plane $z=2000\ \mbox{nm}.$  Top row, from left to right: $f(x,y)$ for a cosine $(\varphi_{\mathrm{CEP}}=0)$ and a sine $(\varphi_{\mathrm{CEP}}=\pi/2)$ pulse, and finally the difference, taking all electrons into account. The bottom row is the same, except it is for electrons only with $E_{kin}>120\ \mbox{eV}.$ In each row the leftmost color map is normalized, the other two were scaled by the same number, i.e., the ratio of the distributions is correct. The parameters are $\mathcal{E}_0=10\ \mbox{GV/m}$ and $a=b=c=100 \ \mbox{nm}$ for all panels.
\label{detect1fig}}
\end{center}
\end{figure}

This fact suggests that when CEP dependent effects are to be observed, special attention has to paid to electrons with high final kinetic energies. Therefore, when investigating the spatial distributions that could be measured by planar detectors that are positioned as shown in Fig.~\ref{geometryfig}, we consider the statistics of high-energy electrons alone. Fig.~\ref{detect1fig} shows the case when the detecting plane is parallel with the polarization direction of the exciting laser pulse. As we can see, sine and cosine pulses produce different distributions already when all electrons are taken into account, but this difference becomes considerably more pronounced, when we focus on electrons with high kinetic energies. Numerically, we have chosen $E_{kin}>120\ \mbox{eV}$ in Fig.~\ref{detect1fig}. For the actual parameter values, CEP-dependence can be observed with a remarkably high contrast in this energy range, thus the effect is expected to be experimentally measurable.

\begin{figure}[htb]
\begin{center}
\includegraphics[width=8cm]{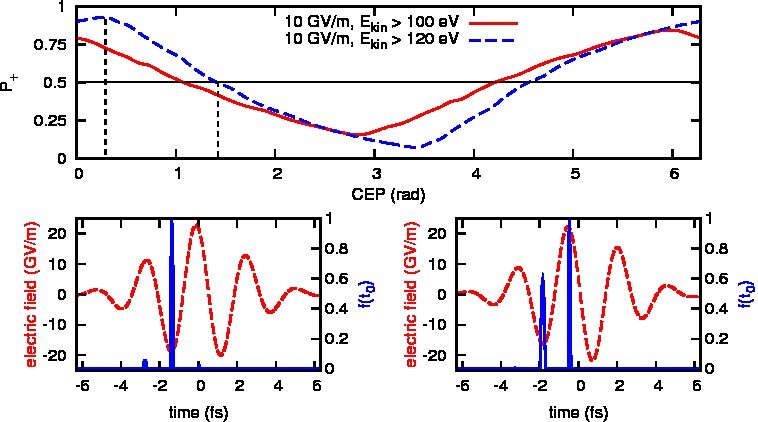}
\caption{Panel (a): CEP dependence of the probability $P_+$ for electrons with high kinetic energies.  The geometrical parameters are  $a=b=c=100 \ \mbox{nm}$ for all curves. Panels (b) and (c): The distribution of the energetic $(E_{kin}>120\ \mbox{eV})$ electrons as a function of $t_0$, i.e., the time instant they leave the surface of the metal (solid blue lines). The $x$ component of the net electric field at $x=a, y=z=0$ is also shown (dashed red line). The corresponding CEP values are indicated by the narrow, dashed  vertical lines in panel (a): $\varphi_{\mathrm{CEP}}=0.1\ \pi$ (b) and $0.42 \ \pi$  (c).
\label{explainfig}}
\end{center}
\end{figure}

In order to analyze the physical origin of CEP dependence, we plot the probability $P_+$ of energetic electrons having positive final momenta $p_x$ as a function of the CEP of the exciting laser pulse. As it is shown by Fig.~\ref{explainfig}a), this function exhibits strong oscillations, the amplitude of which can be as high as 90\% if a suitably high spectral cutoff is applied. The remarkable CEP dependent effects seen in Figs.~\ref{detect1fig}-\ref{explainfig} can be understood by plotting the distribution of the energetic electrons as a function of the time instant they leave the surface. This function, $f(t_0),$ can be seen in Fig.~\ref{explainfig}b) and c) for maximal and zero $P_+$ values, together with the normal component of net electric field at $x=a, y=z=0.$  As we can see, there are a few, narrow time windows in which electrons whose final energy is high, are emitted. This conclusion is obviously not true for resonant nanoparticles (c.f. with waveforms in Fig.~\ref{awayfieldfig}, there are obviously multiple similar time windows in that case), this is why CEP dependence is smeared in the resonant case (Fig.~\ref{spect1fig}). For nonresonant nanoparticles, however, high-energy electrons can be characterized by practically a single $t_0$ value at the $\varphi_{\mathrm{CEP}}$ value where $P_+$ is maximal (see the bottom left panel of Fig.~\ref{explainfig}). The corresponding trajectories are also qualitatively the same: Right after leaving the surface, these electrons start to accelerate in the positive $x$ direction, slow down and turn back when the field changes sign, move towards the surface, get scattered and -- since the electric field changes sign again -- finally accelerate again in the positive $x$ direction, i.e., away from the surface. Therefore, the final momentum of these particles in this direction is positive. On the other hand, when $P_+$ is $1/2$ (bottom right panel of Fig.~\ref{explainfig}), there is an additional family of trajectories that starts at the $x<0$ side of the ellipsoid and after scattering they gain negative final momentum in the $x$ direction. The contributions of the two families cancel, leading to $P_+=P_-=1/2.$ In other words, the bottom row of panels in Fig.~\ref{explainfig} provides a clear physical picture why few-cycle laser pulses impinging on nonresonant nanoparticles can steer photoemitted electrons in a way that strongly depends on their waveform.

\bigskip
{\em Summary}$\ \ $
We developed a model demonstrating strong-field control of plasmonic photoemission
 with few-cycle laser fields.  As a remarkable feature, the control process can be realized at low incident laser intensities exploiting nanoparticle field enhancement. Our results show
 considerable field enhancement and few-cycle plasmon
 oscillations at the same time if the plasmon eigenfrequency is tuned slightly to off-resonance with respect to the laser wavelength.
 By optimizing nanoparticle shape and the time structure of the laser pulse, we
 expect to be able to exercise full control over the motion of electrons in the
 closest nanoscale vicinity of the nanoparticle. This, together with the
 extremely high electron acceleration gradient that are achievable at metal
 nanostructures \cite{DH13,PS13} can enable the
 construction of novel, well-controlled electron sources with nanostructured
 photocathodes for applications in electron injectors \cite{TB08,LT13,PS13b}.
 State-of-the-art electron spectroscopic characterization tools with angular or
 spatial resolution provide ample experimental opportunities to study the predicted
 phenomena. From the fundamental research perspective, plasmonically
 photoemitted electrons can locally probe nanoplasmonic fields, the accurate
 time-resolved mapping of which are expected to shed light on how collective
 electron oscillations build up on the nanoscale within a fraction of a
 femtosecond.

\bigskip
{\em Acknowledgements}$\ \ $
This work was partially supported
by the European Union and the European Social Fund through projects
''Supercomputer, the national virtual lab'' (grant no.: TAMOP-4.2.2.C-11/1/KONV-2012-0010)
and
''Impulse lasers for use in materials science and biophotonics'' (grant no.: TAMOP-4.2.2.A-11/1/KONV-2012-0060),
and by the Hungarian Scientific Research Fund (OTKA) under Contracts No.~T81364 and 109257. P.~D.~acknowledges
 partial support from the Bolyai Fellowship of the Hungarian Academy of
 Sciences and from a Marie Curie Grant of the EU (project "UPNEX", GA302657). We Thank M.~G.~Benedict and M.~Ivanov for useful discussions.

\end{document}